\documentclass[graybox]{svmult}

\usepackage{mathptmx}       % selects Times Roman as basic font
\usepackage{helvet}         % selects Helvetica as sans-serif font
\usepackage{courier}        % selects Courier as typewriter font
\usepackage{type1cm}        % activate if the above 3 fonts are
\usepackage{makeidx}         % allows index generation
\usepackage{graphicx}        % standard LaTeX graphics tool
\usepackage{multicol}        % used for the two-column index
\usepackage[bottom]{footmisc}% places footnotes at page bottom

\newcommand{\R}{{\rm I}\!{\rm R}}

\makeindex             % used for the subject index
\begin{document}

\title*{Inverse Modeling for MEG/EEG data}
\author{Alberto Sorrentino and Michele Piana}
\institute{Alberto Sorrentino \at Dipartimento di Matematica, Universit\`a degli Studi di Genova, Via Dodecaneso 35, Genova, Italy, and CNR--SPIN, Genova, Italy \email{sorrentino@dima.unige.it}
\and Michele Piana \at Dipartimento di Matematica, Universit\`a degli Studi di Genova, Via Dodecaneso 35, Genova, Italy, and CNR--SPIN, Genova, Italy \email{piana@dima.unige.it}}
\maketitle

\abstract*{
%Each chapter should be preceded by an abstract (10--15 lines long) that summarizes the content.
%Please use the 'starred' version of the new Springer \texttt{abstract} command for typesetting the text of the online abstracts (cf. source file of this chapter template \texttt{abstract}) and include them with the source files of your manuscript. Use the plain \texttt{abstract} command if the abstract is also to appear in the printed version of the book.
We provide an overview of the state-of-the-art for mathematical methods that
are used to reconstruct brain activity from neurophysiological data.
After a brief introduction on the mathematics of the forward problem, we discuss standard and recently proposed regularization methods, as well as Monte Carlo techniques for Bayesian inference.
We classify the inverse methods based on the underlying source model, and discuss advantages and disadvantages. Finally we describe an application to the pre--surgical evaluation of epileptic patients.
}

\abstract{
%Each chapter should be preceded by an abstract (10--15 lines long) that summarizes the content. 
We provide an overview of the state-of-the-art for mathematical methods that
are used to reconstruct brain activity from neurophysiological data.
After a brief introduction on the mathematics of the forward problem, we discuss standard and recently proposed regularization methods, as well as Monte Carlo techniques for Bayesian inference.
We classify the inverse methods based on the underlying source model, and discuss advantages and disadvantages. Finally we describe an application to the pre--surgical evaluation of epileptic patients.
}

\section{Introduction}
Neurophysiological investigation based on modalities like electro-- and magneto--encephalography (EEG \cite{niedermeyer2005electroencephalography} and MEG \cite{hamalainen1993magnetoencephalography}), electrocorticography (ECoG, \cite{paetal16}), and stereo EEG (SEEG, \cite{cardinale2013stereoelectroencephalography}) is experiencing an impressive growth in both basic research and clinical applications. The reason of this development is two-fold. First, neurophysiological devices are by far the functional approaches that assure the best possible time resolution (up to 1 millisecond); second, recent hardware developments like multi-channel whole-head EEG and MEG helmets provide a spatial resolution of around 1 cm which, although not as impressive as the time resolution, still permits to draw reliable conclusions from both a scientific and a diagnostic viewpoint. For example, among the available brain imaging modalities, EEG, MEG, and ECoG uniquely allow the realization of functional experiments able to investigate the role of brain oscillatory systems \cite{de2010temporal}. Further, EEG is for sure the gold standard in Brain Computer Interface experiments \cite{wolpaw1991eeg}, while MEG and SEEG are certainly the most promising modalities for the comprehension of  epileptogenic activity and for the neurophysiological interpretation of epileptic seizures \cite{murakami2016correlating}.

Despite these notable potentialities, the application of neurophysiological modalities in research and, in particular, in the clinical workflow, is still limited by the fact that the interpretation of neurophysiological signals is often a very challenging task, which can be accomplished by exploiting powerful and sophisticated computational methods. Methodological problems in this framework are typically of three kinds. First, looking at the data recorded in the sensors' space, these can be encoded by means of classification algorithms based on {\em{machine learning}}, which aims at interpreting the measured times series as different clusters of data mirroring cortical activities of different natures. Second, these same experimental measurements can be modeled in the cortical source space by means of {\em{regularization methods}} that numerically solve the problem of reconstructing the neural currents from the recorded neurophysiological time series. Third, both the measured data and the data-modeling cortical sources can be interpreted in the framework of {\em{connectivity maps}} that can be constructed by designing either statistical or deterministic metrics in both the time and frequency domain. 

The focus of this chapter is the source modeling of cortical sources from the knowledge of electro- or magneto-physiological data. This is a difficult mathematical issue, requiring the solution of a dynamical, highly ill-posed inverse problem. In particular, such ill-posedness implies that the neural configuration explaining the measurements is not unique (there is an infinite number of neural current distributions producing the same dataset, \cite{dassios2013definite}) and this technical difficulty has inspired the adoption of many different strategies (some proposed by the inverse problems community, some others coming from the engineering framework) for the selection of the optimal  neural constellation from the infinite set of possible solutions. The available algorithms are usually divided into two classes, based on the physical model used to represent the neural currents: \textit{distributed} methods assume a continuous current distribution and solve a linear inverse problem consisting in recovering the dynamics of the local strength of the current density at each point of a computational grid introduced in the brain volume; \textit{dipolar} methods introduce in the reconstruction procedure the information that the neural sources can be modeled as a set of a small number of point-like currents (current dipoles), whose parameters (position, orientation, strength) have to be recovered; assuming this dipolar model, the inverse problem is non linear since the measurements have a strongly
non-linear dependence with respect to the unknown source locations.
Distributed methods address the MEG data analysis as an image restoration problem whereby the restored map solves a constrained minimization; among these methods we recall Tikhonov regularization with L2 norm \cite{hamalainen1994interpreting} and L1 norm \cite{uutela1999visualization} on the penalty term; various type of beamformers \cite{sekihara2001reconstructing, barnes2003statistical}, which spatially filters the signal to focus the source as a weighted combination of the measurements. On the other hand dipolar methods estimate dipoles parameters by non linear optimization
techniques; among these we recall the dipole fitting procedure, the multiple signal classification (MUSIC) \cite{mosher1998recursive}, and its evolution called recursively applied and
projected MUSIC \cite{mosher1999source}; genetic algorithms \cite{uutela1998global}, which find the sources with a trial-and-error procedure. However, both classes suffer from well-known
shortcomings: among the distributed methods, the L2-norm estimation tends to provide solutions which are too wide-spread from a neurophysiological viewpoint,
L1-norm estimation requires an expert user interpretation of the reconstructed sources and beamformers suppress temporally correlated sources; among dipolar approaches, many non-linear optimization methods suffer from convergence problems; furthermore many methods consider a fixed (pre-determined or estimated from
the data) number of sources, maintaining their position and often also orientation across time.

Recently, owing to the increase in the available computational power, Bayesian methods have become feasible \cite{somersalo2003non,caetal08,sorrentino2009dynamical,long2011state,miao2013efficient}. They cast the inverse problem as a problem of
statistical inference by means of Bayesian statistics: the unknown and the measurements are modelled as Random Variables (RV) and the solution of the inverse
problem is the posterior probability density function of the unknown obtained by the Bayes theorem. Bayesian methods typically use the whole time series as input data and therefore can be considered as methods that address the dynamical source modeling problem. However, some of these approaches aim at reconstructing the source configuration at a given time point, while in others the sequential application of Bayes theorem, which requires a prior density at each time step, is mediated by the use of the Chapman-Kolmogorov equation. The main strengths of Bayesian approaches are in their notable generality (they can be applied in very general frameworks with minimal a priori assumptions) and in the fact that they can naturally account for a priori information coming from either physiology or experimental measurements provided by other modalities. On the other hand, these methods are typically demanding from a computational viewpoint and require the (often difficult) optimization of several input parameters. 

The next sections will provide an overview of the state-of-the-art and some open issues concerned with the source modeling problem in neurophysiology. Specifically, Section 2 describes the forward modeling of data formation for the electro- and magneto-physiology problems; Section 3 will discuss several inversion approaches to data modeling and Section 4 will show an application to the case of measurements acquired during epileptic seizures. Finally, our conclusions will be offered in Section 5.

\section{Data formation}
\label{sec:data}
The mathematical framework that allows the definition of the forward problem for EEG, MEG, ECoG and SEEG is given by the Maxwell equations under the quasi-static condition, i.e.
\begin{equation}\label{eq:maxwell-1}
\nabla \cdot E = \frac{\rho}{\epsilon_0}~~,
\end{equation}
\begin{equation}\label{eq:maxwell-2}
\nabla \times E = 0~~,
\end{equation}
\begin{equation}\label{eq:maxwell-3}
\nabla \cdot B = 0~~,
\end{equation}
and
\begin{equation}\label{eq:maxwell-4}
\nabla \times B = \mu_0 J~~,
\end{equation}
where $E = E(r,t)$ and $B=B(r,t)$ are the electric and magnetic fields at position $r \in \R^3$ and time $t$, respectively; $\epsilon_0$ and $\mu_0$ are the electric and magnetic permittivities in vacuum, respectively, and $J = J(r,t)$ is the current density. Equation (\ref{eq:maxwell-2}) allows the introduction of the scalar potential $V=V(r,t)$ such that
\begin{equation}\label{eq:scalar-potential}
E = - \nabla V~~,
\end{equation}
and equation (\ref{eq:maxwell-4}) implies that
\begin{equation}\label{eq:density-1}
\nabla \cdot J = 0~.
\end{equation}

In each point of the brain volume, the current density $J$ is made by the superposition of two contributions, i.e.
\begin{equation}\label{eq:density-2}
J = J_p + J_v~,
\end{equation}
where the primary current $J_p$ passing through the axons induces a passive volume current density $J_v$, which solves the local Ohm law
\begin{equation}\label{eq:density-3}
J_v(r,t) = \sigma (r) E(r,t)~,
\end{equation}
where $\sigma(r)$ is the tissue conductivity at $r$. Combining equations (\ref{eq:scalar-potential})-(\ref{eq:density-3}) leads to
\begin{equation}\label{eq:EEG-forward-problem}
\nabla \cdot J_p(r,t) = -\nabla \cdot (\sigma(r) \nabla V(r,t))~,
\end{equation}
which defines the forward model for the formation of EEG, SEEG, and ECoG signals. In fact, assuming that the conductivity is known (for example, thanks to the information provided by some anatomical medical imaging modality like Magnetic Resonance Imaging) and given a vector field $J_p = J_p(r,t)$, the numerical solution of equation (\ref{eq:EEG-forward-problem}) at sampled points $r_1,\ldots,r_N$ on the skull provides the EEG signal recorded by the helmet sensors placed at those points and at time $t$ \cite{puetal11}. On the other hand, if the points are on either a limited portion of the cortex or a line through the brain, solving the equation provides the signal recorded by the sensors of an ECoG grid and a SEEG device, respectively. 

The definition of the forward model mimicking the MEG signal requires the introduction of the vector potential $A=A(r)$ such that
\begin{equation}\label{eq:vector-potential-1}
B = - \nabla \times A~,
\end{equation}
whose existence is a consequence of equation (\ref{eq:maxwell-3}). In order to constrain the many possible shapes of the field $A$ satisfying (\ref{eq:vector-potential-1}), the Coulomb gauge condition
\begin{equation}\label{eq:coulomb}
\nabla \cdot A = 0
\end{equation}
is imposed in the following. Under such condition, the vector potential solves the Poisson equation
\begin{equation}\label{eq:poisson}
\nabla^2 A = \mu_0 J~,
\end{equation}
whose solution can be written as
\begin{equation}\label{eq:solution-poisson}
A(r) = \mu_0 \int_{\R^3} G(r,r^{\prime}) J(r^{\prime}) dr^{\prime}~,
\end{equation}
given the Green function
\begin{equation}\label{eq:green}
G(r,r^{\prime}) = \frac{1}{4\pi|r-r^{\prime}|}~.
\end{equation}
Replacing (\ref{eq:solution-poisson})-(\ref{eq:green}) into (\ref{eq:vector-potential-1}) and exploiting the fact that
\begin{equation}\label{eq:formula}
\nabla \times \frac{J(r^{\prime})}{|r-r^{\prime}} = J(r^{\prime}) \times \frac{r-r^{\prime}}{|r-r^{\prime}|^3}
\end{equation}
leads to the Biot-Savart equation
\begin{equation}\label{eq:biot-savart}
B(r,t) = \frac{\mu_0}{4\pi} \int_{\R^3} [J_p(r^{\prime},t) -\sigma(r^{\prime}) \nabla V(r^{\prime},t)] \times \frac{r-r^{\prime}}{|r-r^{\prime}|} dr^{\prime}~.
\end{equation}
The Biot-Savart equation is combined with equation (\ref{eq:EEG-forward-problem}) to define the forward model for MEG. In fact, assuming once again that the conductivity is known, and given the vector field $J_p = J_p (r,t)$, the computation of $V(r,t)$ in (\ref{eq:EEG-forward-problem}) for $r$ everywhere in the cortex and then of $B(r,t)$ in (\ref{eq:biot-savart}) at specific points $r$ outside outside (but close) the scalp provides the signal recorded by the MEG sensors, placed at those points.

\section{The inverse problem}
\label{sec:IP}

Spatio--temporal localization of brain activity from MEG/EEG data requires to solve a so--called \textit{inverse problem}: given the recorded time series, compute an estimate of the underlying neural currents. In a discrete setting, we will be calling the recorded data $y_{1:T}:=\{y_t\}_{t=1}^T$, where $y_t$ is a vector whose components represent the data recorded by the sensors. The spatio--temporal evolution of the neural current is represented by $x_{1:T}$, where $x_t$ is a vector that contains a discrete representation of the neural current; more details on such discrete representation will be given shortly.
We assume that the two processes are related by
\begin{equation}
	y_t = f(x_t) + n_t \hspace{1cm} t=1,\dots,T~~~,
	\label{eq:ip}
\end{equation}
where $n_t$ is the noise at time $t$ and the \textit{forward model} $f(\cdot)$ is not further specified at this stage, because its actual form depends on the specific choice of the discrete representation of the neural current.

It is well known that the MEG/EEG inverse problem is ill--posed; in particular, it has no unique solution, because there exist current distributions that do not produce any magnetic field. 
In regularization theory, this is usually dealt with by minimizing a functional that combines the fit with the data and some penalty term on the unknown, so as to encourage some desired property of the solution(smoothness, sparsity, and so on).
\begin{equation}
	\hat{x}_{1:T} = \arg \min \left( \| y_{1:T} - f(x_{1:T})\|^2_2 + \lambda P(x_{1:T}) \right)
	\label{eq:regularization}
\end{equation}

In the Bayesian framework, ill--posedness is dealt with by changing the point of view: rather than trying to solve the inverse problem, one aims at quantifying the amount of information that is available on the unknown underlying neural currents. The information is coded in probability distributions: specifically, one is interested in computing or approximating the posterior distribution $p(x_{1:T}|y_{1:T})$, i.e. the probability distribution of the unknown, conditioned on the measured data; this distribution is notoriously given by Bayes theorem:
\begin{equation}
	p(x_{1:T}|y_{1:T}) \propto p(y_{1:T}|x_{1:T}) p(x_{1:T})
	\label{eq:bayes}
\end{equation}
where $p(y_{1:T}|x_{1:T})$ is the \textit{likelihood} function and $p(x_{1:T})$ is the \textit{prior} distribution. 

Notably, there is a clear connection between these two approaches: under Gaussian noise assumptions, the likelihood function is
\begin{equation}
	p(y_{1:T}|x_{1:T}) \propto \exp \left( - \| y_{1:T} - f(x_{1:T})\|^2_2 \right)
\label{eq:bayes_reg_1}
\end{equation}
and the regularized solution described by (\ref{eq:regularization}) can be interpreted as the maximum of the log--posterior described by (\ref{eq:bayes}) when the prior is of the form 

\begin{equation}
p(x_{1:T}) \propto \exp(-P(x_{1:T}))
\label{eq:bayes_reg_2}
\end{equation}
However, Bayesian methods typically aim at approximating/calculating the whole posterior distribution, rather than just its maximum, in order to assess the uncertainty of the estimate. As one can easily understand, the whole posterior distribution (\ref{eq:bayes}) lives in a very high--dimensional space, which makes it almost hopeless to compute it or approximate it. As we will see, Bayesian methods will try to simplify the problem in different ways.

\subsection{Classification of inverse methods}

The specific choice of the discrete representation of the neural current will lead to a first classification of inverse methods.

Indeed, one can assume that the neural current is a continuous vector field in the whole brain volume; for computational purposes, such continuous vector field can be discretized by choosing an appropriate grid of points $\{r_i\}_{i=1}^N$ with $N$ large enough. With reference to the notation of the previous Section, the discrete representation of the neural current is then such that $(x_t)_i = J^p(r_i,t)$. In this case, we will refer to as the \textit{distributed} source model, the inverse problem can be written as a linear problem
\begin{equation}
	y_t = \sum_{i=1}^N G(r_i) \cdot (x_t)_i + n_t = {\bf G} \, x_t + n_t
\label{inverse:distributed}
\end{equation}
where $\textbf{G}$ is usually called \textit{leadfield} matrix. In this case, $N$ is usually a relatively large number ($\sim 10,000$) of voxels.

Alternatively, one can assume that the neural current is a finite set of point--like sources, termed \textit{current dipoles}: the number of dipoles is not known a priori, neither their locations, orientations and strengths. Here, $x_t$ is a finite set of dipoles:
$x_t = (r_t^1,q_t^1,\dots,r_t^N,q_t^N)$. In this case, we still have
\begin{equation}
	y_t = \sum_{i=1}^{N_t} G(r_t^i) \cdot q_t^i + n_t 
\label{inverse:dipoles}
\end{equation}
but this does not reduce to a linear problem, because the source locations are unknown. Typical 
dipole models contain a relatively small ($< 10$) number of dipoles.\\

A second type of classification concerns the way different inverse methods treat the temporal dependence of the data.
Indeed, due to the high temporal resolution of MEG/EEG data (around $1,000$ Hertz), the data tend to be fairly smooth, and the underlying electrical currents are likely to exhibit a similar degree of smoothness in the temporal variable.
In fact, most methods developed until the mid 2000s solved the inverse problem independently at different time points; this is mainly due to the limited computational resources available at that time, that made the solution of the full spatio--temporal unfeasible; we will refer to these methods as \textit{static} inverse methods. More recently developed methods try to model the temporal dependence in order to reduce the ill--posedness further; while this remains impractical in some cases (see e.g. the network of High Performance Computers needed to run a Kalman filter \cite{long2006large}), it is becoming increasingly doable. We will refer to these methods as \textit{dynamic} methods.

\subsection{Methods for the distributed model}

We start with a short review of static methods, that are still the most widely used ones thanks to their computational efficiency.

\paragraph{Minimum Norm Estimate (MNE)}

The Miminum Norm Estimate \cite{hamalainen1994interpreting} (MNE) corresponds to solving a simplified version of (\ref{eq:regularization}), where each time point is treated independently and with $P(x_t) = \| x_t \|^2_2$; in Bayesian terms, this amounts to calculating the MAP of $p(x_t|y_t)$. The advantage of this approach is that analytical calculation is feasible, thanks to the relatively small size of the matrices involved:
\begin{equation}
	\hat{x}^{MNE}_t = ( G^T G + \lambda I)^{-1} G^T y_t 
	\label{eq:wmne}
\end{equation}
On the other hand, it is well known that this choice has two main drawbacks: first, it tends to provide overly smooth estimates of the neural current; second, it tends to create a bias towards superficial sources, because superficial sources are closer to the sensors and therefore weaker sources, encouraged by the $\ell^2$ norm, produce stronger signals. To overcome this last point, a \textit{weighted} MNE variant has been proposed in \cite{lin2006assessing}, in which the penalty term is designed to compensate for the depth bias $P(x_t) = x_t^T R^{-1} x_t$.

To overcome the first point \cite{uutela1999visualization} proposes an $\ell^1$ penalty term $P(x_t) = \|x_t\|_1$. A recent Bayesian alternative has been described in \cite{calvetti2015hierarchical}, where a Bayesian hierarchical model is used to code a priori anatomical information, and a hyper--parameter tunes the degree of sparsity of the solution.

\paragraph{Mixed norm estimates}

In order to account for the temporal smoothness of MEG/EEG data, in the last decade several regularization methods have been proposed \cite{ou2009distributed,tian2012two,gramfort2012mixed} that use a penalty with a mixed norm \cite{kowalski2009sparse}. An example of these mixed norms is the $\ell_{1,2}$ norm that is defined as follows

\begin{equation}
	\| x_{1:T} \|_{1,2} = \sum_{i=1}^N \sqrt{ \sum_{t=1}^T (x_t^i)^2 }
\end{equation}
By using $P(x) = \|x\|_{1,2}$ one obtains a convex optimization problem; in this case, the analytical solution is not available, but the convexity of the functional guarantees uniqueness of the solution, and convergence results for a class of iterative algorithms with proximal operators \cite{parikh2014proximal}. This type of mixed norm methods provide pleasant solutions with few active regions that exhibit a smooth temporal behaviour. On the other hand, they are slightly tricky to use, as the iterative algorithms used to compute the solution require some fine--tuning of the parameters. 

\paragraph{Kalman filtering}

In a Bayesian setting, it is natural to include prior information on the temporal behaviour of the neural sources by means of a Kalman filter. Rather than computing the whole posterior distribution (\ref{eq:bayes}), the Kalman filter provides a mean to compute the so called \textit{filtering} distributions, i.e. $p(x_t|y_{1:t})$, for $t=1,\dots,T$. Indeed, by assuming that the neural currents and the data form a  Hidden Markov Model \cite{rabiner1986introduction}, the two following ``Bayesian filtering'' equations provide a formal way to calculate the filtering distributions:

\begin{eqnarray}
p(x_t|y_{1:t}) = \frac{p(y_t|x_t)p(x_t|y_{1:t-1})}{p(y_t|y_{1:t-1})}~~~, \label{eq:bf_bayes} \\
p(x_{t+1}|y_{1:t}) = \int p(x_{t+1}|x_t) p(x_t|y_{1:t}) dx_t~~~.
\label{eq:bf_kolmogorov}  
\end{eqnarray}
For a linear Gaussian model, all the filtering distributions are Gaussian, and are therefore completely characterized by their mean and covariance matrix. If we denote by $\bar{x}_{t|s}$ and $\Gamma_{t|s}$ the mean and covariance matrix, respectively, of $p(x_t|y_{1:s})$, then the Kalman filter is given by:

\begin{equation}
\label{filtro4}
\bar{x}_{t|t} = \bar{x}_{t|t-1} + K_{t} (y_t - G \bar{x}_{t|t-1})
\end{equation}
\begin{equation}
\label{filtro5}
\Gamma_{t|t}=(I-K_{t} G) \Gamma_{t|t-1 } 
\end{equation}
where
\begin{equation}
\label{filtro3}
K_{t}=\Gamma_{t|t-1}G^T (G \Gamma_{t|t-1} G^T + \Sigma)^{-1}
\end{equation}

\begin{equation}
\label{pred1}
\bar{x}_{t+1|t}=F \bar{x}_{t|t}
\end{equation}
\begin{equation}
\label{pred2}
\Gamma_{t+1|t}=F\Gamma_{t|t}F^T+ \Delta
\end{equation}

However, these formulas are computationally demanding, due to the size of the involved matrices; in particular, for a brain discretization containing $10,000$ points, the covariance matrix of an unconstrained model is a square matrix of size $30,000 \times 30,000$.
For this reason, in \cite{long2011state} a network of high performance computers is used to calculate the full problem; in order to reduce the computational cost, in \cite{galka2004solution,lamus2012spatiotemporal} the full Kalman filter is decomposed in a large number of small--dimensional problems by considering only short--range interactions.

\subsection{Methods for the dipolar model}

Methods for estimating multiple current dipoles have to face two main difficulties: first, the number of dipoles is unknown, which translates into a model order selection problem; second, the problem is genuinely non--linear, implying that the corresponding functional to be minimized has local minima, and deterministic algorithms are doomed to failure.

\paragraph{Global optimization methods}

For years, dipole fitting methods have relied on subjective choices of initialization and non--linear optimization of individual dipole parameters with Levenberg--Marquardt. In \cite{huang1998multi}, a multi--start algorithm is proposed that tried to avoid subjectivity of the results, by several random initialization of the optimization algorithm.
In \cite{uutela1998global} the authors describe the use of simulated annealing and of genetic algorithms for optimizing a regularized functional. In both cases, the number of dipoles has to be fixed in advance.

\paragraph{Bayesian Monte Carlo methods for static dipoles}

In the Bayesian setting, one can formally include the number of dipoles among the unknowns, and try and make inference on it by exploring variable--dimension models with reversible jumps \cite{green1995reversible}.
In order to make the problem computationally tractable, it is common to assume that the number of dipoles and their locations do not change in the time window under investigation; this means that the posterior distribution of interest is
\begin{equation}
	p(N,r^1,\dots,r^N, q_1^1, \dots, q_t^1, \dots, q_T^N, \dots, q_T^N|y_{1:T})
\end{equation}
Monte Carlo sampling on this space is still prohibitive but, thanks to the linear relationship between the data and the dipole moments, one can marginalize out (i.e., treat analytically) the dipole moments and do Monte Carlo sampling only for $p(N,r^1,\dots,r^N)$.
In \cite{jun2005spatiotemporal} the authors assume a uniform prior for the number of dipoles, and use reversible--jump Markov Chain Monte Carlo (MCMC) to approximate the posterior distribution. In \cite{sorrentino2014bayesian,sommariva2014sequential}, the authors assume a Poisson prior for the number of dipoles, and use sequential Monte Carlo (SMC) samplers \cite{del2006sequential} to approximate the posterior distribution; as SMC samplers employ multiple Markov Chains running in parallel, they are less likely to remain trapped in local maxima.

\paragraph{Bayesian Monte Carlo methods for dynamic dipoles}

Starting from \cite{somersalo2003non}, Bayesian inference for dynamic dipoles has been described in several studies \cite{sorrentino2009dynamical,sorrentino2013dynamic,miao2013efficient,antelis2013dynamo,chen2015bayesian,vivaldi2016bayesian}.
Like for the distributed case, the easiest thing to do is to use the Bayesian filtering recursion described by equations (\ref{eq:bf_bayes}) -- (\ref{eq:bf_kolmogorov}). However, in this case the analytic solution is not available, because the data depend non--linearly on the dipole locations and on the number of dipoles. Therefore one can use \textit{particle} methods \cite{doucet2009tutorial} to approximate the filtering distributions. The idea is to propagate in time a set of Monte Carlo samples (particles) by importance sampling and Markov Chain Monte Carlo, possibly combined. As time goes by, the sample set accumulates information on the underlying sources. By their nature, filtering algorithms tend to provide a poor localization of dipolar sources at their appearance, just because there is little information accumulated from the past. For this reason, in \cite{vivaldi2016bayesian} a particle \textit{smoothing} algorithm is proposed, that approximates the conditional distributions $p(x_t|y_{1:T})$ for $t=1,\dots,T$.

\section{An application to epilepsy}

Thanks to their outstanding temporal resolution, MEG and EEG provide a unique window into the human brain, with many applications ranging from basic neuroscience to the clinical use. Here we provide some insight into a clinical application: the pre--surgical evaluation of epileptic patients. Indeed, this is the most straightforward clinical application of the source localization problem.

Epilepsy is a chronic disease that affects about fifty million people worldwide, with 30\% of cases being refractory to medication. Focal epilepsy is the most common type of epilepsy in adults \cite{nguyen2013prevalence}, where epileptic seizures are generated in a relatively small area of the brain, referred to as \textit{epileptogenic zone} (EZ). When focal epilepsy is refractory to medication, surgical ablation of the EZ is considered as a possible solution. However, localization of the EZ is often not straightforward: current clinical practice envisages the use of intra--cerebral recordings \cite{cardinale2013stereoelectroencephalography,andrzejak2015localization}.
The use of MEG/EEG for localization of the EZ still needs further validation, but is increasingly considered; in particular, recent studies suggest that MEG may be capable of estimating not only the location, but also the size of the EZ \cite{bouet2012towards}.

Here we consider, as an example, the localization of an EZ from 32--channel EEG recordings. Data come from the example BESA database (BESA, Munich): 164 spikes were recorded from an epileptic patient and averaged, using the peak of the spike as a trigger.
The EEG signals were sampled at 320 Hz and filtered with a Butterworth forward high-pass filter with cut-off frequency of 5 Hz. The head model is a three-layer model including the brain, the skull and the scalp; while there is no cerebrospinal fluid (CSF) in the model, the effect of the CSF is partly accounted for by assuming an anisotropic skull conductivity. The tangential conductivity within the skull is modeled to be three times larger than the radial conductivity across the skull.

\begin{figure}
	\includegraphics[width=6cm]{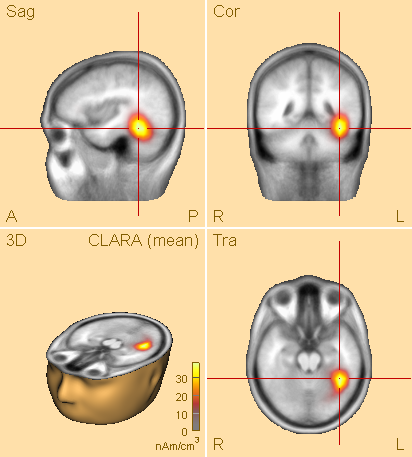}
	\hspace{.2cm}
	\includegraphics[width=6cm]{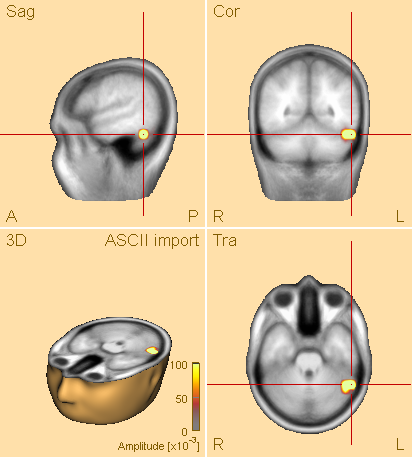}
	\caption{Reconstructions obtained for an epileptic patient in the proximity of the spike. Left: CLARA. Right: smoothing algorithm.}
	\label{fig:epilepsy}
\end{figure}

In Figure \ref{fig:epilepsy} we report the results obtained by applying two different methods: CLARA (Classical LORETA Analysis Recursively Applied), which is a distributed method with a Laplacian penalty term, and depth--weighting; a particle smoothing algorithm, which is used to approximate the smoothing distribution for a dynamic multi--dipole model.
The Figure reports the reconstructed map at a particular time, in correspondence with the peak of the spike. We notice that the two figures have substantially different interpretations: while the images produced by CLARA represent the estimated strength of the neural current, the images produced by the smoothing algorithm represent the posterior probability of there being a dipole at any particular location. Despite of these differences, the two methods appear to agree pleasantly on the putative location of the EZ. 

\section{Conclusions}

Even though EEG has been around for almost a century, and MEG for almost half of it, inverse modeling of MEG/EEG data remains a challenging research field, due to the inherent ill--posedness of the problem.
After the very first localization attempts in the 80s, the search for better spatial priors in the 90s, the inclusion of the temporal dimension in the 2000s, most likely the present decade will be remembered for the prevalence of connectivity studies, and for the transfer to the clinical practice. Concerning connectivity studies, a lot of effort is currently spent in understanding the dynamics of the connectivity networks. On the other hand, the transfer to clinical practice is requiring a substantial validation work, using both simulations and experimental data. Finally, a fascinating future perspective concerns the use of these devices for brain computer interfaces.
For sure we are only at the beginning of exciting developments in many different fields, including thought--controlled devices and improved clinical treatment of epilepsy.

%\begin{acknowledgement}
%If you want to include acknowledgments of assistance and the like at the end of an individual chapter please use the \verb|acknowledgement| environment -- it will automatically render Springer's preferred layout.
%\end{acknowledgement}

\bibliography{biblio}
\end{document}